\documentclass[aps,prb,preprint,endfloats]{revtex4}

\usepackage{graphicx}

\newcommand{\Qv}{\vec{Q}_{/\!/}} 
\newcommand{\Qs}{Q^2_{/\!/}} 
\newcommand{\Q}{Q_{/\!/}} 

\begin{document}

\title{Spin Waves in Ultrathin Ferromagnets: Intermediate Wave Vectors}

\author{A. T. Costa}
\affiliation{
Depto. de Ci\^encias Exatas, 
Universidade Federal de Lavras,
37200-000 Lavras, Minas Gerais, Brazil}

\author{R. B. Muniz}
\affiliation{
Instituto de F\'isica,
Universidade Federal Fluminense,
24210-340 Niter\'oi, Rio de Janeiro, Brazil}

\author{D. L. Mills}
\affiliation{
Department of Physics and Astronomy,
University of California,
Irvine, California 92697 U. S. A.}

\begin{abstract}
Our earlier papers explore the nature of large wave vector spin waves
in ultrathin ferromagnets, and also the properties and damping of spin
waves of zero wave vector, at the center of the two dimensional
Brillouin zone, with application to FMR studies. The present paper
explores the behavior of spin waves in such films at intermediate wave
vectors, which connect the two regimes. For the case of Fe films on
Au(100), we study the wave vector dependence of the linewidth of the
lowest frequency mode, to find that it contains a term which varies as
the fourth power of the wave vector. It is argued that this behavior
is expected quite generally. We also explore the nature of the
eigenvectors of the two lowest lying modes of the film, as a function
of wave vector. Interestingly, as wave vector increases, the lowest
mode localizes onto the interface between the film and the substrate,
while the second mode evolves into a surface spin wave, localized on
the outer layer. We infer similar behavior for a Co film on Cu(100),
though this evolution occurs at rather larger wave vectors where, as
we have shown previously, the modes are heavily damped with the
consequence that identification of distinct eigenmodes is
problematical.
\end{abstract}

\maketitle

\section{Introduction}

The nature of spin motions in ultrathin ferromagnetic films and more
generally in magnetic nanostructures is a fundamental topic, important
also from the point of view of applications. It is the motion of
magnetic moments within ultrathin films that allow giant
magnetoresistive read heads respond to the bits on hard discs, and
more recently in the elements incorporated into prototype magnetic
random access memory (MRAM). Thus, issues such as the frequency of
precession of magnetic moments and the damping mechanisms operative in
magnetic nanostructures is a topic of interest from the perspective
not only of fundamental physics, but from the point of view of
contemporary device technology as well. As entities incorporated into
devices become smaller and smaller, the physics of spin motions with
spatial gradients emerges as a central issue. We see this from the
discussions of the injection of spin polarized currents into ultrathin
ferromagnets through use of both point contact devices and spatially
resolved optical excitation \cite{1}. Useful insight into the influence of
finite wave vector effects on the frequencies, character of spin wave
eigenvectors, and damping may be obtained through the study of finite
wave vector spin excitations in ultrathin ferromagnetic films of
infinite extent.

We have been engaged in theoretical studies of the nature of spin
waves in ultrathin ferromagnets, both free standing and adsorbed on
metal substrates, along with their damping. These calculations are
based on use of an itinerant electron description of the
ferromagnetism in the film, and a realistic electronic structure of
the the film/substrate combination. Details of our approach may be
found in our study of the Fe[110] monolayer on W[110] \cite{2}. We
initially concentrated our efforts on the systematic features of these
modes, throughout the appropriate two dimensional Brillouin zone \cite{2,3,4,5}. 
A focus was placed on the strong intrinsic damping of these
modes, which increases dramatically as one moves from the center of
the Brillouin zone out to large wave vectors. A striking prediction
emerged from these studies. Very near the zone center, the lowest
lying acoustical spin wave mode has a very long lifetime, whereas even
the first standing spin wave mode suffers substantial damping. For an
N layer film, with $N$ in the vicinity of six or eight, the higher
frequency standing waves are so heavily damped they are barely
perceptible, if at all, in our calculated spectral functions. As one
moves out into the Brillouin zone, the damping becomes so severe that
one is left only with a single broad feature in the spin fluctuation
spectrum, whose peak displays dispersion expected of a spin wave
mode. This picture contrasts dramatically with that provided by a
Heisenberg model of a film with localized magnetic moments coupled
together by exchange interactions. In such a picture, for each wave
vector in the two dimensional Brillouin zone, one has $N$ spin wave
modes, and each mode has infinite lifetime. A discussion of the nature
of the modes of such a Heisenberg film has been given by one of the
authors some decades ago \cite{6} and we refer the reader to a review
article which covers early theoretical studies of spin waves at the
surface of Heisenberg magnets, and in films \cite{7}.  In regard to
ultrathin metallic ferromagnets on metallic substrates, we now have in
hand beautiful spin polarized electron loss data which confirm our
predictions regarding the nature of the spin wave modes in such
systems \cite{8}. We obtain an excellent quantitative account of both the
dispersion with wave vector of the single, heavily damped feature
found in the experiments, as well as its width and asymmetric
lineshape \cite{5}.

The damping mechanism operative in these analyses is, from the point
of view of many body physics, a magnetic analogue to the well known
Landau damping process responsible for the heavy damping experienced
by plasmons in simple metals.  One sets up a coherent collective mode
of the plasma at time $t=0$ (the plasmon of metal physics), and the
amplitude of the collective mode decays with time as its energy is
transferred to incoherent particle hole pairs. Interestingly, for the
ideal, collisionless plasma, this transfer does not involve energy
dissipation, and is reversible. Thus, one may observe plasma echos in
the presence of Landau damping \cite{9}. In the case of itinerant
ferromagnets, the spin wave is a collective oscillation whose nature
is very similar to the plasmon in a metal, when viewed in the
framework of many body physics \cite{10}. This mode may also decay to
particle hole pairs, very much as in the plasmon case. In the
ferromagnet, conservation of spin angular momentum in the decay
process requires the particle hole pairs to be spin triplet
excitations. These are commonly referred to as Stoner excitations in
the literature on itinerant ferromagnetism \cite{9}. We pause to remark
that an intriguing possibility which emerges from this analogy is that
of observing FMR echoes which would be the FMR analogue of the plasma
echos just discussed. An ideal sample would be an ultrathin
ferromagnet deposited on a non magnetic substrate of finite thickness,
with thickness small compared to the spin diffusion length. An example
of such a system would be a Ag film grown on GaAs (100), followed by
an ultrathin Fe film which is possibly capped with Ag.

When an ultrathin ferromagnet is adsorbed on a metallic substrate, in
our studies we find that the decay process is far more efficient than
in a bulk ferromagnet with the same atomic constituents. Hence one
finds very short lifetimes for the large wave vector modes. The decay
of the collective spin wave to the Stoner excitation spectrum of the
film/substrate combination leads to a spin current which transports
angular momentum from the ferromagnetic film into the substrate, hence
leading to a decrease in the transverse magnetization associated with
the spin motion in the ferromagnet. This same mechanism has been
discussed and analyzed in the literature on the ferromagnetic
resonance linewidths; in essence one is discussing the lifetime of the
acoustic spin wave mode of zero wave vector. In the FMR literature,
this is referred to as the “spin pumping” contribution to the
linewidth. Spin pumping was proposed as an important source of FMR
linewidths in ultrathin films by Berger and Slonczewski, in seminal
papers \cite{11}.  These authors employed a simple model description of a
film of localized, Heisenberg like moments coupled to a bath of
conduction electrons. Spin pumping was observed experimentally by
Woltersdorf and his colleagues \cite{12} and many others since that
time. Subsequent theoretical studies provided a very good account of
the data \cite{13}.  In a recent paper, we have explored the predictions
which follow from our approach to spin wave damping.  For Fe films
grown on a Au(100) substrate, we obtain an excellent quantitative
account of the data in ref. \onlinecite{14}. We also obtain a very fine
theoretical description of systematics of the linewidth found in
trilayers grown on the Cu (100) surface \cite{15}. Thus, our method appears
to provide a very satisfactory description of intrinsic linewidths
observed in SPEELS studies of large wave vector spin waves in the
Co/Cu(100) system, and also the linewidth found for the zero wave
vector spin wave in FMR studies of Fe on Au(100).

The present paper presents studies which address the connection
between the very small (zero, essentially) wave vector modes studied
in FMR and the large wave vector regime addressed in the SPEELS study
of the Co/Cu(100) system. Two issues are addressed here. First, in
regard to the spin pumping contribution to the linewidth, we explore
its wave vector dependence.  We also examine the nature of the
eigenvectors of the low lying spin waves in the film as one moves away
from the zone center into the Brillouin zone. Here we find behavior
for the Fe/Au(100) system that is very striking. The two lowest lying
spin wave modes evolve into waves in which one (the lowest) localizes
on the interface between the substrate and the film as wave vector
increases, whereas the second mode localizes on the outer surface of
the film.  We argue that a very similar picture applies to the very
different Co/Cu(100) system as well, though the physics is obscured by
the heavier damping found in the latter system, in the relevant regime
of wave vector. If, then, one wishes to interpret the SPEELS data in
terms of a simple picture of spin wave modes in the film, the SPEELS
loss spectrum would receive its dominant contribution not from the
lowest lying mode in the film, as proposed by the authors of ref. \onlinecite{8},
but rather the second mode in the hierarchy of spin wave modes of the
film. We note that we set forth this proposal in our earlier
publication \cite{5}, and the results presented in this paper reinforce
this interpretation. The comparison we make between these two
ultrathin films of different crystal structure suggests in our mind
that the behavior we find may be expected rather generally in the
ultrathin ferromagnets.

The outline of this paper is as follows. In section II we provide the
reader with a brief summary of the early theoretical analyses of the
nature of low lying spin wave modes in Heisenberg films. These
studies, very relevant to our present discussion, are not so well
known in the present era so a reminder of the concepts which follow
from these papers will provide a setting for what emerges from the
work reported here. We then present our results in section III, and
concluding remarks are found in section IV.

\section{Surface Spin Waves on Heisenberg Ferromagnets}

We begin by considering an ideal semi infinite Heisenberg
ferromagnet. The term ideal here describes a model system in which the
strength of the underlying exchange interactions in and near the
surface assume the same values as they do in the bulk material. In
such a case, one may envision forming two semi infinite crystals by
beginning with an infinitely extended crystal, and then cutting all
exchange bonds which cross a mathematical plane between two atomic
planes.

Through study of one particular model surface, Wallis and co workers
\cite{16} pointed out that such an ideal surface can support surface spin
waves, with amplitude that decays exponentially as one moves into the
crystal from the surface. The properties of the surface wave found by
these authors differ strikingly from those of the more familiar
Rayleigh surface phonons which propagate on crystal surfaces. In the
bulk of the material, and in the long wavelength limit we know well
that the frequency of spin waves varies quadratically with wave
vector. We have $\hbar\omega(\vec{Q})=DQ^2$ in cubic crystals. 
The surface spin wave studied in
ref. \onlinecite{16} exists for all wave vectors in the surface Brillouin
zone. If one considers the surface spin wave with two dimensional wave
vector $\Qv$, its frequency lies below that of the manifold of bulk spin
waves whose wave vector projection onto the plane of the surface
assumes the value $\Qv$. However, in the long wavelength limit the
frequency of the surface spin wave is $\hbar\omega_S(\Qv)=D\Qs$ , with D the bulk exchange
stiffness. The “binding energy” of the surface spin wave in the
long wavelength limit arises from the terms quartic in the wave
vector. In contrast, in the long wavelength limit, the Rayleigh
surface phonon propagates with velocity less that of any bulk phonon
with the same wave vector parallel to the surface, so the surface wave
and bulk wave dispersion curves differ to leading order in $\Qv$. As noted
earlier, the amplitude of the surface spin wave discussed in ref. \onlinecite{16}
decays to zero exponentially as one penetrates into the bulk
material. If one describes this by the expression $\exp[-\alpha(\Qv)l_z]$ where $l_z$
labels an atomic plane, one finds $\alpha(\Qv)$ is proportional to $\Qs$ in the long wavelength
limit, whereas for the Rayleigh surface phonon, the decay constant is
linear in $\Q$. While the authors of ref. \onlinecite{16} studied one specific model
of an ideal Heisenberg ferromagnet surface, subsequent discussions
showed that the features just discussed are robust and follow for a
very wide range of models of the surface, including those in which
exchange interactions near the surface differ substantially from those
in the bulk \cite{6,7}. For the ideal surface, the criterion for the
existence of surface spin waves is as follows \cite{6}. When the surface is
formed by the bond cutting procedure described earlier, one must cut
exchange bonds which are non-normal to the surface.

While we are not concerned in the present paper with the nature of
thermal spin fluctuations at finite temperature in our model film,
interesting issues arise when one discusses the near surface behavior
of thermal spin fluctuations. The surface spin waves are eigenmodes of
the Heisenberg Hamiltonian and thus are present as thermally excited
spin waves. However, if one calculates the amplitude of thermal
fluctuations in the magnetization near the surface, in the low
temperature limit the contribution from the surface spin waves is
exactly and precisely cancelled by a deficit in the density of bulk
spin waves which results from the formation of the surface wave; there
is a “hole” in the density of bulk spin wave modes that leads to
this cancellation. This was demonstrated first for the model studied
in ref. \onlinecite{16}, and later shown to be robust \cite{6,17} and insensitive
to the microscopic details of the surface environment. In the end, one
finds that amplitude of the thermal fluctuations in the surface to be
twice that deep in the bulk of the material, and one may derive an
analytic expression for the dependence of the mean spin deviation as a
function of distance into the material from the surface \cite{6,17}, in
the limit of low temperatures.

We now turn our attention to films. If we have an isolated film with
two identical surfaces, then of course all the spin wave eigenvectors
must have well defined parity under reflection through the midpoint of
the film. Suppose we consider a wave vector $\Qv$ in the surface Brillouin
zone sufficiently large that the quantity $\alpha(\Qv)$ introduced above 
satisfies $\alpha(\Qv)N>1$,
with $N$ the number of atomic layers within the film. Then the two
lowest lying modes of the film with wave vector $\Qv$ will have the
character of surface spin waves. One mode (that with the highest
frequency, for reasons to be given below) will be odd under reflection
through the film with a displacement pattern that decays exponentially
as one moves into the interior of the film from either surface, and
the other will be even parity also with displacements localized near
the surfaces. The frequency splitting between the two modes will be
proportional to $\exp[-2\alpha(\Qv)]$.

Now suppose, for the film just discussed, we let $\Qv\rightarrow 0$.  
As we proceed with this limit, we enter the regime where $\alpha(\Qv)N<1$, 
and the two surface mode
eigenvectors must continuously and smoothly evolve into the two lowest
lying $\Qv=0$ modes of the finite film. The lowest of these has zero frequency
(in the absence of an applied Zeeman field), and is the uniform mode
of the film, wherein all spins precess rigidly and in phase. This is
an even parity mode. The next highest mode is an odd parity standing
wave mode whose eigenvector vanishes at the midpoint of the film. The
perpendicular wave vector assumes the value $\pi/N$ as $\Qv\rightarrow 0$, 
so this mode has
finite, non zero frequency. The odd parity mode thus has higher
frequency than the even parity mode, and one thus expects the odd
parity mode to have the higher frequency throughout the surface
Brillouin zone, though of course in principle one may have film
parameters in which a crossing of the dispersion curves occurs.

These comments provide us with a setting for the results we shall
present below which, of course, are based on a fully itinerant
electron description of the spin waves in our film. One important
difference between the film we consider here, and the Heisenberg film
with two identical surfaces is that in our case the two surfaces are
inequivalent. One is the interface between the film and the substrate
upon which it is absorbed, and the second is the outer surface, with
vacuum above. We shall see that at the center of the Brillouin zone,
we find the lowest mode to be the uniform mode, whose eigenvector is
modified however, by the enhanced moment at the film vacuum interface.
The next highest mode looks very much like the first odd parity
standing wave. As the wave vector increases, rather than realize the
even and odd parity surface waves, we shall see that the lower mode
localizes near the film/substrate interface, whereas the second mode
(not strictly odd parity for our film), localizes on the outer
surface. In our earlier study \cite{5} of the Co film on Cu (100), we used
adiabatic theory to calculate effective Heisenberg exchange
interactions between various nearest and next nearest neighbor
moments. As one sees from Table II, at the outer surface, and in the
layer against the substrate the nearest neighbor exchange interactions
are enhanced substantially over the values deep in the film, with the
effective exchange on the outer surface larger than that at the
interface between the substrate. Such enhanced effective exchange
plays an important role in “binding” the spin waves to the
surface, and to the film/substrate interface.

\section{Properties of Intermediate Wave Vector Spin Waves for Fe/Au(100)
and for Co/Cu(100)}

As remarked above, we have used the programs developed for our
analysis of the spin pumping linewidtth for the Fe film on Au(100) to
explore the nature of the spin waves in the Fe film and their damping
near the center of the Brillouin zone. We begin by presenting these
results, and then we turn to our considerations for the Co film on
Cu(100). The first mentioned system was used in the experimental
studies of the spin pumping contribution to the FMR linewidth by Urban
and collaborators \cite{12}, and large wave vector spin waves in the second
system were explored in the SPEELS experiments reported by Vollmer and
co workers \cite{8}.

The results below were extracted from studies of the wave vector and
frequency dependent susceptibility $\chi_{+,-}(\Qv,\Omega;l,l')$ of 
four and eight layer Fe(100)
films adsorbed on the Au(100) surface.  From this response function,
we form the spectral density function 
$A(\Qv,\Omega;l_\perp)=\frac{1}{\pi}\chi_{+,-}(\Qv,\Omega;l_\perp,l_\perp)$. 
From the spectral density
function we may extract the dispersion relation of the spin wave modes
by following the trajectory with wave vector of the resonant peaks in
this response function, and the linewidth is obtained from the width
of these structures. Information on the nature of the eigenvectors may
be obtained from the transverse susceptibility itself through a
procedure described below. We refer the reader to section II of
ref. \onlinecite{14} for a discussion of the physical significance of these two
quantities.

\begin{figure}
\centerline{\includegraphics[width=0.8\textwidth,clip]{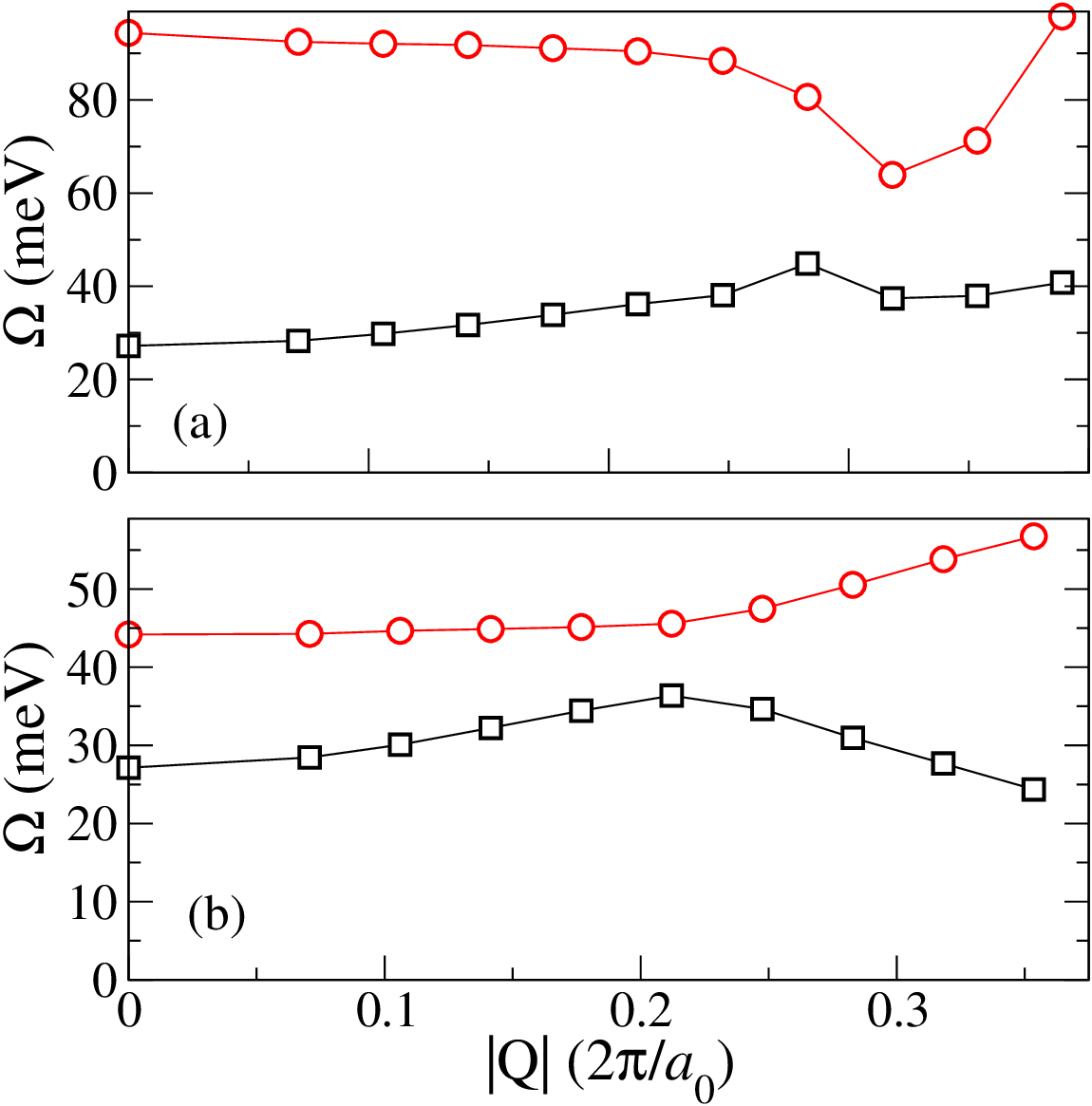}}
\caption{For the (a) four layer, and (b) the eight layer Fe film on
Au(100), we show the dispersion relation of the two lowest lying modes
of the film as a function of reduced wave vector, along the [11]
direction in the Brillouin zone. As discussed in the text, a Zeeman
field has been imposed so the lowest mode has finite frequency at the
zone center.}
\label{fig1}
\end{figure}

When we plot the dispersion curves of the two lowest lying spin wave
modes for the Fe film near the center of the Brillouin zone, we find a
most interesting level crossing, as illustrated in Fig. \ref{fig1}. In
Fig. \ref{fig1}(a), we show the dispersion relation of these two modes for the
four layer Fe(100) film on Au(100), for wave vectors along the [11]
direction. With increasing wave vector, the lowest lying mode shows
positive curvature, while the first standing wave mode shows negative
curvature. The two dispersion curves start to cross, and we see a
hybridization gap. We note that the hybridization gap is a feature
present by virtue of the fact that the two surfaces of the film are
distinctly different; at the outer surface of the film, we have an
interface with the vacuum, and the inner surface is the interface with
the substrate. In a free standing film, the low frequency mode would
have even parity, the first standing wave would have odd parity, and
symmetry would then prohibit the mixing that leads to the hybrization
gap. We remark that, as in our earlier studies of the spin pumping
contribution to the linewidth, we have added a Zeeman field to our
Hamiltonian which renders the frequency of the lowest frequency mode
finite in the limit of zero wave vector. The field we use is
unphysically large, but as discussed earlier \cite{14} so long as the spin
wave frequencies are small compared to the energy scale of the one
electron band structure, all the field does is to shift all modes
upward in frequency by the Zeeman energy. There are computational
advantages to introducing this shift. Linewidths are linear in
frequency at zero wave vector, as we have shown earlier \cite{14} so when
we plot the ratio of the linewidth to the frequency of the mode, we
have a ratio independent of applied field so long as the spin wave
frequencies are low. In Fig. \ref{fig1}(b) we show dispersion curves for the
eight layer film. We see, as expected, the exchange contribution to
the frequency of the first standing wave mode is quite accurately four
times less than that found for the four layer film.

\begin{figure}
\centerline{\includegraphics[width=0.8\textwidth,clip]{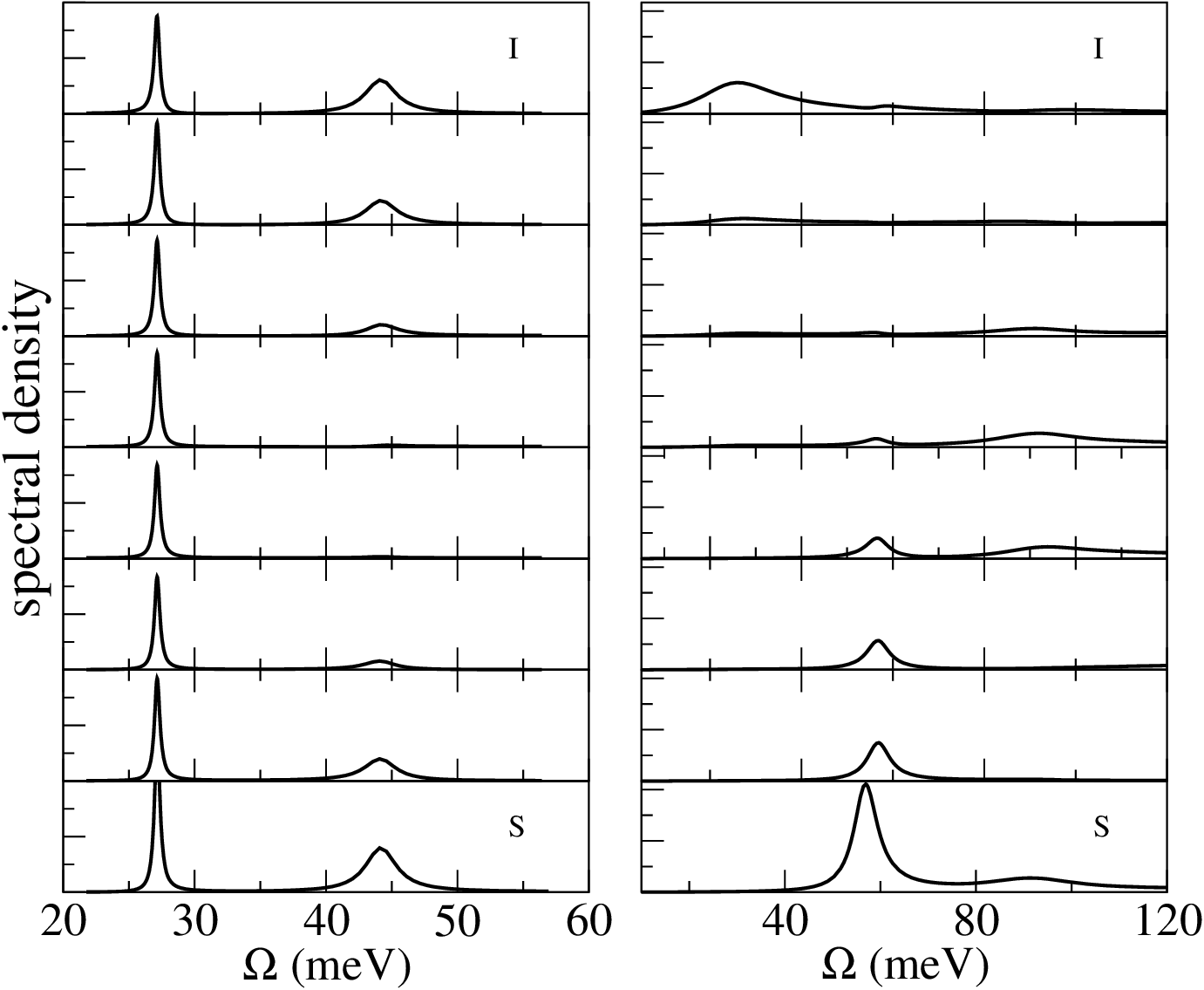}}
\caption{We plot the frequency variation of the spectral density 
$A(\Qv,\Omega;l_\perp)$ as a function of layer index $l_\perp$, 
for the eight layer Fe film on Au(100) and
for two selected wave vectors. In the left panel, we have zero wave
vector, and in the right panel the reduced wave vector is 0.25 along
the [11] direction. The top entries, labeled I, is the layer adjacent
to the Au substrate, and the lowest entry labeled S is the outer
surface layer of the film.}
\label{fig2}
\end{figure}

In Fig. \ref{fig2}, for the eight layer film we plot the spectral density
function $A(\Qv,\Omega;l_\perp)$ as a function of frequency, for various layers in the
film. The layer labeled I at the top of each plot is the interface
between the film and the substrate, while the lowest layer labeled S
is the outer surface layer. The leftmost panel shows the spectral
densities for zero wave vector, whereas the right most panel gives
these for a reduced wave vector of 0.25. As explained earlier, \cite{14}
the integrated intensity of each peak can be interpreted as the square
of the eigenvector of the mode associated with the peak. In the
leftmost panel, we see that the low frequency mode is indeed the
uniform mode of the film. It is the case that the amplitude in the
surface is somewhat larger than in the inner layers. This is an effect
with origin in the enhanced surface moment at the surface/vacuum
interface. The higher frequency mode is clearly a classical standing
wave, with an eigenvector when squared has a cosine squared variation
layer number, and a node in the center of the film. The wavelength
perpendicular to the film is twice the film thickness.

As wave vector increases, we we see from the rightmost panel in
Fig. \ref{fig2}, the lower mode becomes a localized spin wave mode, with
eigenvector localized at the film/substrate interface. The high
frequency mode evolves into a surface spin wave, localized on the
outer surface. Interestingly, the higher frequency mode is narrower
than the low frequency mode. Our previous studies suggest that the
broadening at fixed wave vector parallel to the surface increases with
the gradient in the eigenvector in the direction normal to the
surface. The low frequency mode is considerably more localized than
its higher frequency partner, which suggests it should indeed be
broader.

\begin{figure}
\centerline{\includegraphics[width=0.9\textwidth,clip]{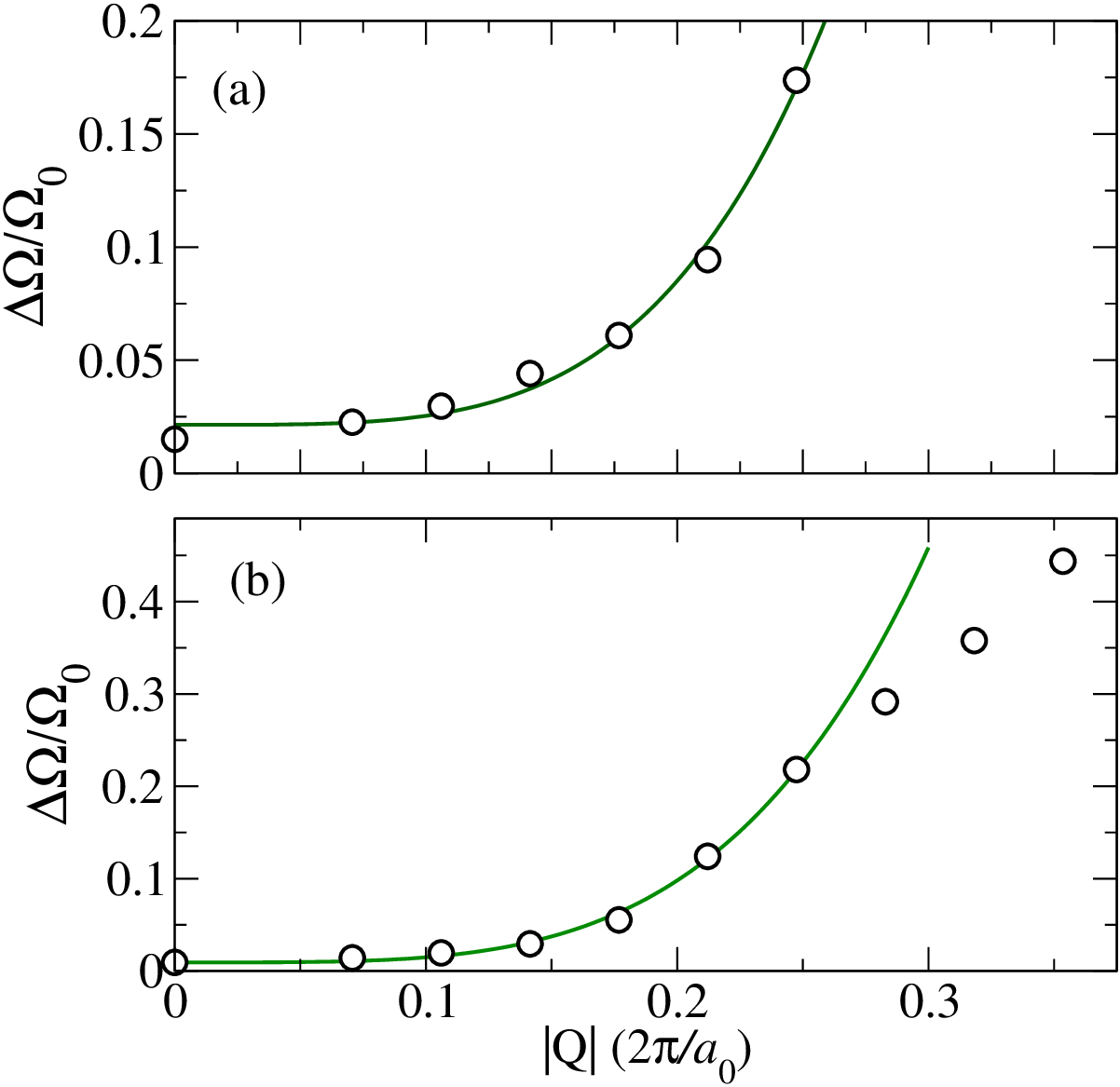}}
\caption{Linewidths of the lowest mode as a function of wavevector along the [11]
direction in a 4 layer (a) and a 8 layer (b) Fe
film on Au(100). The solid curves are fittings to $Q^4$ functions.}
\label{fig_linewidth}
\end{figure}

In Fig. \ref{fig_linewidth}, we in (a) we show the wave vector dependence of the
linewidth divided by the mode frequency for the four layer film, and
in (b) we show this for the eight layer film. At zero wave vector, we
have the spin pumping contribution to the linewidth we \cite{14} and others
\cite{11,13} have discussed earlier. This falls off inversely with the
thickness of the ferromagnetic film. The solid line in these figures
assumes that the wave vector dependent portion of the linewidth scales
as $\Q^4$, and we see this fits the data very well indeed over a rather wide
range of wave vectors near the center of the Brilluoin zone.

A simple argument shows that the linewidth must vary as the fourth
power of the wave vector, as we find numerically. We illustrate this
for a very simple case, spin waves in an infinitely extended
ferromagnet as described by the random phase approximation applied to
the one band Hubbard model. It is our view that this conclusion
applied to our much more complex system as well, but the formal
analysis will be very involved. We remark that the argument presented
below is applicable to multi band descriptions of the spin waves in
the bulk, if the Lowde Windsor paramterization \cite{19} of the on site
Coulomb interaction is employed.

For this simple case, the dynamic susceptibility has the well know form
\begin{equation}
\chi_{+,-}(\vec{Q},\Omega) = \frac{\chi^{(0)}_{+,-}(\vec{Q},\Omega)}
{1+U\chi^{(0)}_{+,-}(\vec{Q},\Omega)}\, .
\label{eq1}
\end{equation}

If we write $\chi^{(0)}_{+,-}(\vec{Q},\Omega)=\chi^{(0)R}_{+,-}(\vec{Q},\Omega)+
i\chi^{(0)I}_{+,-}(\vec{Q},\Omega)$, then the spectral density function which 
contains the spin wave signature is
\begin{equation}
A(\vec{Q},\Omega)=\frac{1}{\pi}\frac{\chi^{(0)I}_{+,-}(\vec{Q},\Omega)}
{\left[ 1+U\chi^{(0)R}_{+,-}(\vec{Q},\Omega)\right]^2+
\left[U\chi^{(0)I}_{+,-}(\vec{Q},\Omega)\right]^2}\, .
\label{eq2}
\end{equation}

In the long wavelength limit, the quantity $1+U\chi^{(0)R}_{+,-}(\vec{Q},\Omega)$    
has a zero at the spin wave frequency $\Omega(\vec{Q})=DQ^2$. We then expand  
$\chi^{(0)R}_{+,-}(\vec{Q},\Omega)$  as follows:
\begin{equation}
\chi^{(0)R}_{+,-}(\vec{Q},\Omega) =\chi^{(0)R}_{+,-}(\vec{Q},\Omega(\vec{Q}))+
\dot{\chi}^{(0)R}_{+,-}(\vec{Q},\Omega(\vec{Q})) \left[\Omega-\Omega(\vec{Q})\right]+
\cdots
\label{eq3}
\end{equation}

The imaginary part of $\chi^{(0)}_{+,-}(\vec{Q},\Omega)$ vanishes at zero frequency, 
and this function is linear in frequency for small frequencies. Hence in the limit of
small wave vector, we may write $\chi^{(0)I}_{+,-}(\vec{Q},\Omega(\vec{Q}))=
\chi^{(0)I}_{+,-}(\vec{Q},DQ^2)=DQ^2\dot{\chi}^{(0)I}_{+,-}(\vec{Q},0)$.  
It is simple to show from the explicit expression for 
$\chi^{(0)}_{+,-}(\vec{Q},\Omega)$ that $\dot{\chi}^{(0)I}_{+,-}(0,0)$ vanishes, 
and also that $\dot{\chi}^{(0)I}_{+,-}(\vec{Q},0)$ is an even
function of wave vector. Hence, for small values of the wave vector
one may write $\dot{\chi}^{(0)}_{+,-}\approx bQ^2$ so that we have 
$\chi^{(0)I}_{+,-}(\vec{Q},\Omega(\vec{Q}))\approx bDQ^4$. We may then, 
in the low frequency long wavelength limit make the replacement
$\dot{\chi}^{(0)R}_{+,-}(\vec{Q},\Omega(\vec{Q}))\approx 
\dot{\chi}^{(0)}_{+,-}(0,0)$ in Eq. \ref{eq3}. The spin wave
density can then be written, in the long wavelength low frequency
limit as
\begin{equation}
A(\vec{Q},\Omega) = \frac{m}{\pi}
\frac{\gamma Q^4}{\left[\Omega-DQ^2\right]^2 + \left[\gamma Q^4\right]^2}\, .
\label{eq4}
\end{equation}
Here $m=n_\uparrow - n_\downarrow$, $\gamma=mU^2bD$, and we have used 
$\dot{\chi}_{+,-}^{(0)R}=-1/U^2m $.  

It follows from Eq. \ref{eq4} that for the models
of bulk spin waves encompassed by the above discussion the linewidth
scales as the fourth power of $Q$ very much as we find for our numerical
studies in the ultrathin films. In the film calculations, we obtain a
finite linewidth at zero wave vector by virtue of the applied Zeeman
field. Even if a Zeeman field is applied to the bulk, at zero wave
vector the linewidth of the uniform mode must vanish by virtue of the
Goldstone theorem applied to the Hamiltonian, which is form invariant
under spin rotations. As we have argued earlier \cite{14}, in the ultrathin
films, the breakdown of translational symmetry normal to the surface
allows a finite linewidth when the wave vector parallel to the surface
is zero; the mode is not a uniform spin precession in the entire
system of the film substrate combination.

The spectral density plots shown in Fig. \ref{fig2} provide information on the
variation of the square of the eigenvector of a particular mode, as
one scans through the layer number in the film. We have devised a
means for extracting the eigenvector itself from the function 
$ \chi_{+,-}(\Qv,\Omega;l_\perp,l'_\perp) $. We first
evaluate, for a selected value of $\Qv$, the response function at the
frequency $\Omega_\alpha(\Qv)$ of the mode of interest. This frequency 
is chosen to be the frequency of a selected peak in a spectral density 
plot such as those in Fig. \ref{fig2}. Then the eigenvector is generated 
from the eigenvalue problem
\begin{equation}
\sum_{l'_\perp}\chi_{+,-}(\Qv,\Omega;l_\perp,l'_\perp)e_\alpha(\Qv;l'_\perp)=
\lambda(\Qv)e_\alpha(\Qv;l_\perp)\, .
\label{eq5}
\end{equation}
The eigenvectors generated by this scheme are in general complex, with
an amplitude and a phase.

\begin{figure}
\centerline{\includegraphics[width=0.7\textwidth,clip]{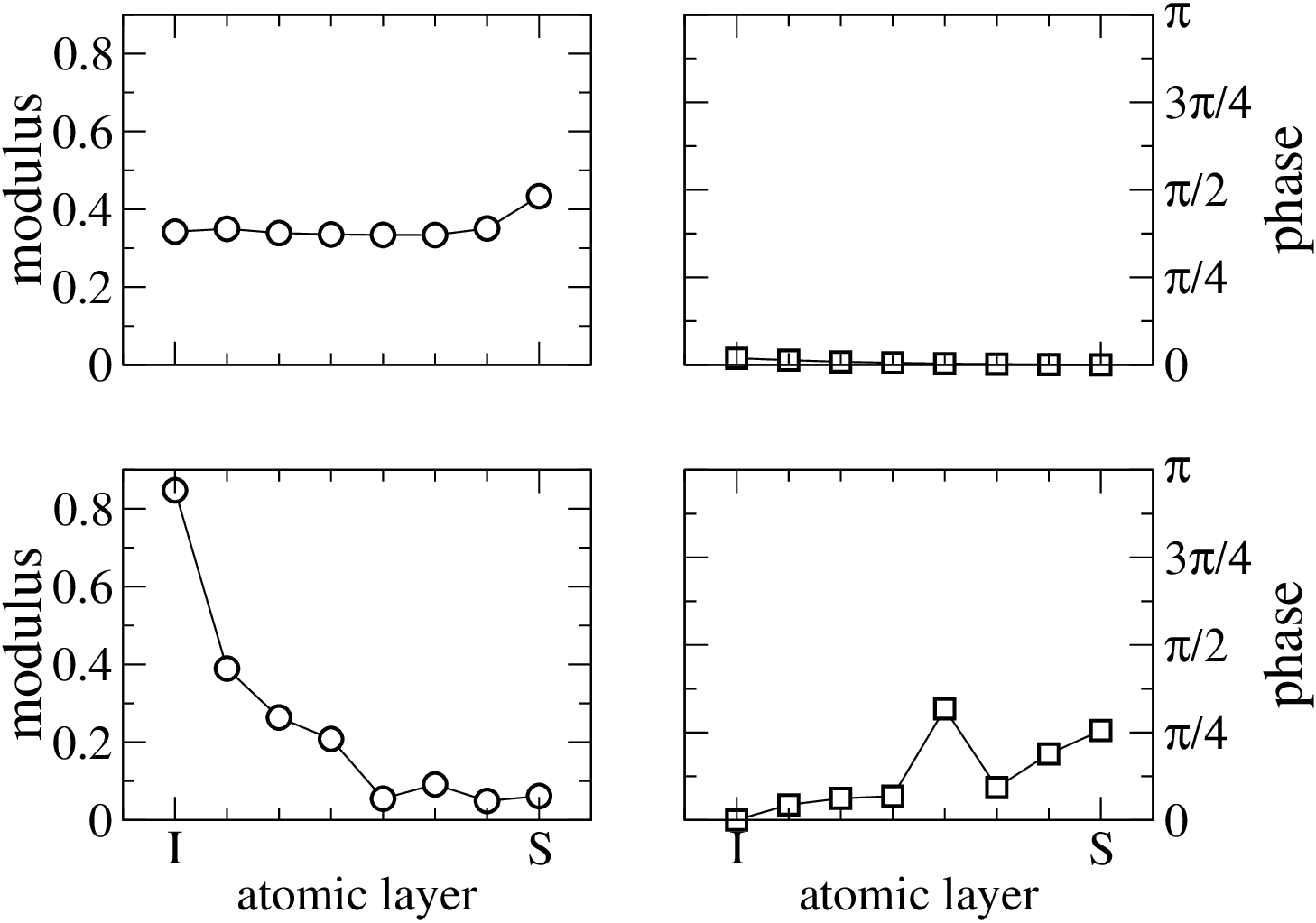}}
\caption{The amplitude and phase of the eigenvector associated with
the lowest frequency mode of the eight layer Fe film on Au(100). The
top two panels show the eigenvector of the mode at $\Qv=0$, and the bottom
two panels give the same for a reduced wave vector of 0.25 along the
[11] direction in the two dimensional Brillouin zone. The layer
labeled I is the Fe layer against the substrate, and the layer labeled
S is the outermost surface layer of the film.}
\label{fig3}
\end{figure}

In Fig. \ref{fig3}, we show the amplitude and phase of the eigenvector
associated with the lowest mode in the eight layer Fe film on Au. The
top two panels show the modulus and phase of the eigenvector at the
center of the two dimensional Brillouin zone. We see the mode is
indeed nearly uniform across the film. The increased amplitude in the
surface layer has its origin in the fact that in the outer layer, the
moment is larger than it is in the middle of the film. Notice the
phase is zero across the film, so the various layers precess in phase,
as expected from simple phenomenology. The lower two panels show the
amplitude and phase of the mode at the reduced wave vector of 0.25 in
the [11] direction. The leftmost layer labeled I is the interface with
the substrate, and the rightmost layer is the outer surface layer. We
see the mode is quite localized near the interface.

\begin{figure}
\centerline{\includegraphics[width=0.7\textwidth,clip]{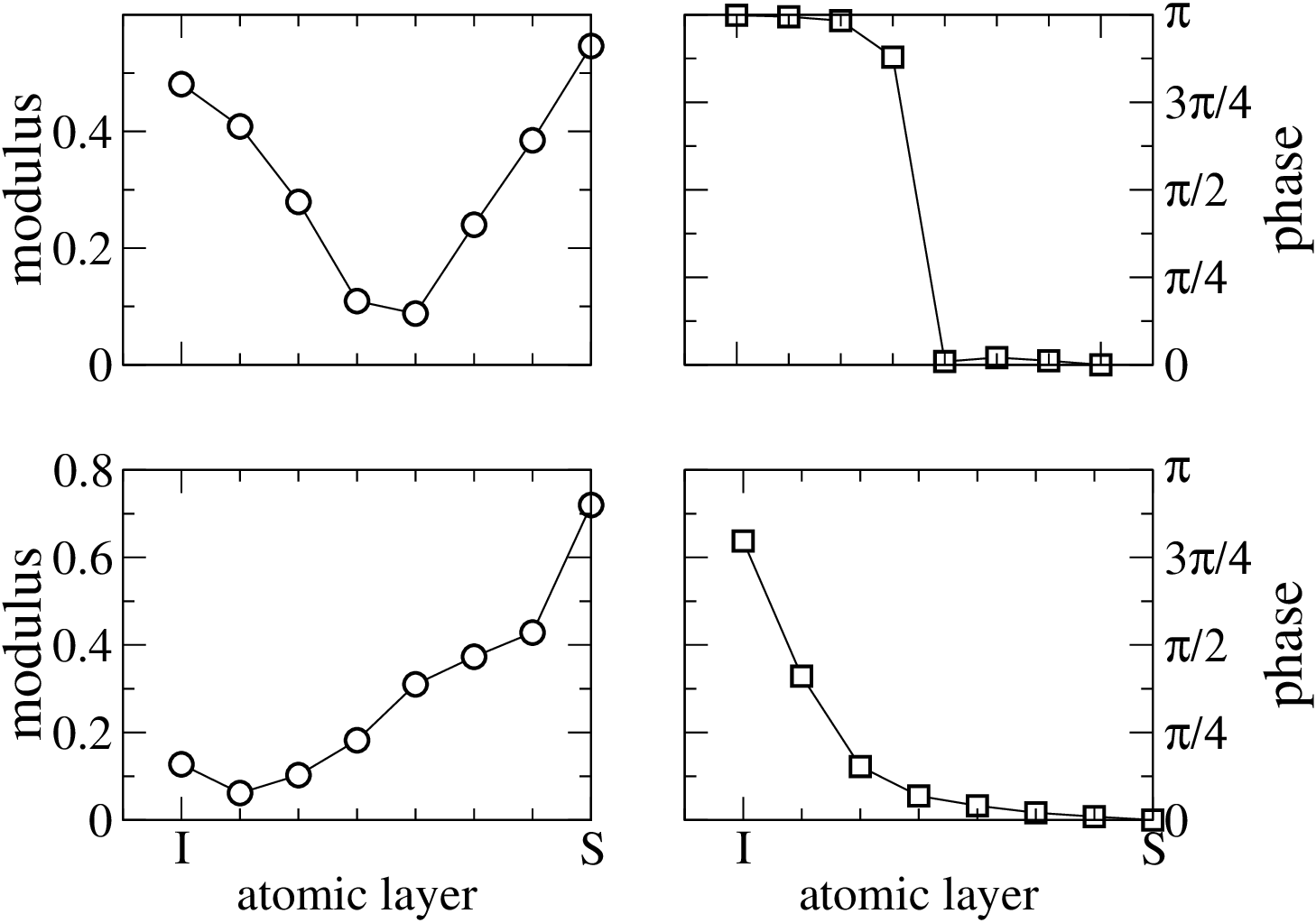}}
\caption{The same as Fig. \ref{fig3}, but now for the second mode of the film.}
\label{fig4}
\end{figure}

We show the same information for the second mode of the film in
Fig. \ref{fig4}. At the center of the Brillouin zone, the amplitude and phase
information show a mode whose profile is rather closely described by
the simple standing wave pattern $ \cos[\pi(l_\perp-1)/7] $ expected 
for the lowest standing wave
mode in the film, with one half wavelength trapped between the
surfaces. The profile is distorted a bit from this form by the an
enhancement of the amplitude in the outer surface layer, as in the
uniform mode. By the time the reduced wave vector is 0.25 (two lower
panels), we see that the mode is localized on the outer surface.

We next inquire if similar behavior is found for the Co film on
Cu(100) which we have studied earlier. When we explore this issue
within the full dynamical theory used above, we see very similar
trends. However, the hybridization between the two lowest spin wave
modes and also their tendency to localize at the interface with the
substrate or the surface appears to occur somewhat farther out in the
Brillouin zone. In this region, the damping has become sufficiently
severe that we have not been able to extract clear eigenvectors for
the two modes utilizing the method we have employed for the case of Fe
on Au(100).  To examine this question for this sytem, we have resorted
to calculations based on the Heisenberg model for this film. In our
previous publication, in Table II we provide values for the exchange
interactions between all nearest and next nearest neighbors, for the
eight layer Co film on Cu(100). Interestingly, we see strong
enhancement of the exchange interactions between nearest neighbors in
both the surface layer, and also those within the layer closest to the
interface.

\begin{figure}
\centerline{\includegraphics[width=0.8\textwidth,clip]{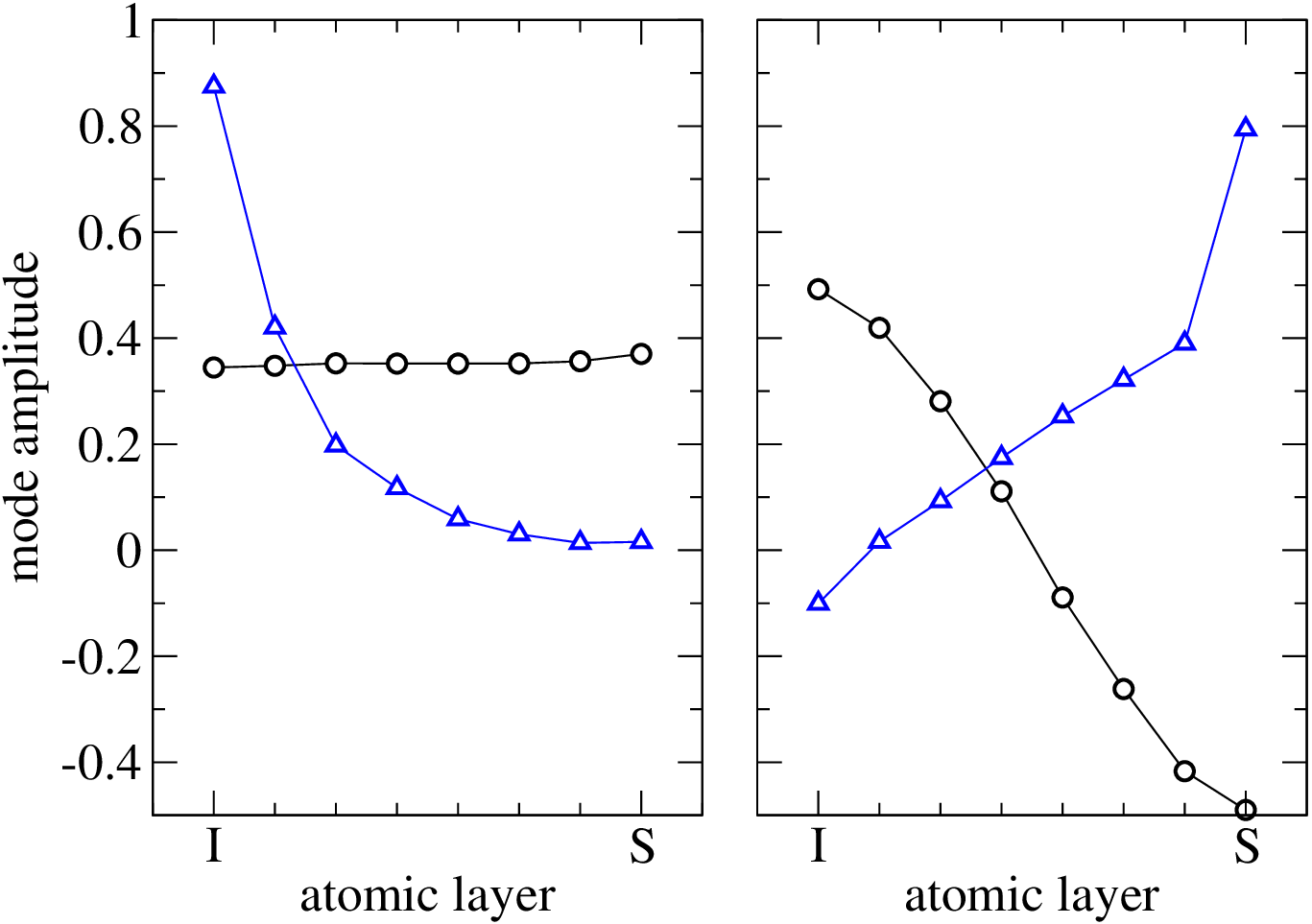}}
\caption{We show eigenvectors for Co on Cu(100) calculated using the
Heisenberg model, as described in the text. The left panel shows the
behavior of the lowest mode at the center of the Brilluoin zone (open
circles), and at the reduced wave vector of 0.6 along the [11]
direction in the Brillouin zone. (triangles). The right panel gives
the same information for the second mode. The layer labeled I is the
layer of Co spins adjacent to the Cu(100) substrate, and the layer
labeled S is the outer surface layer.}
\label{fig5}
\end{figure}

\begin{figure}
\centerline{\includegraphics[width=0.8\textwidth,clip]{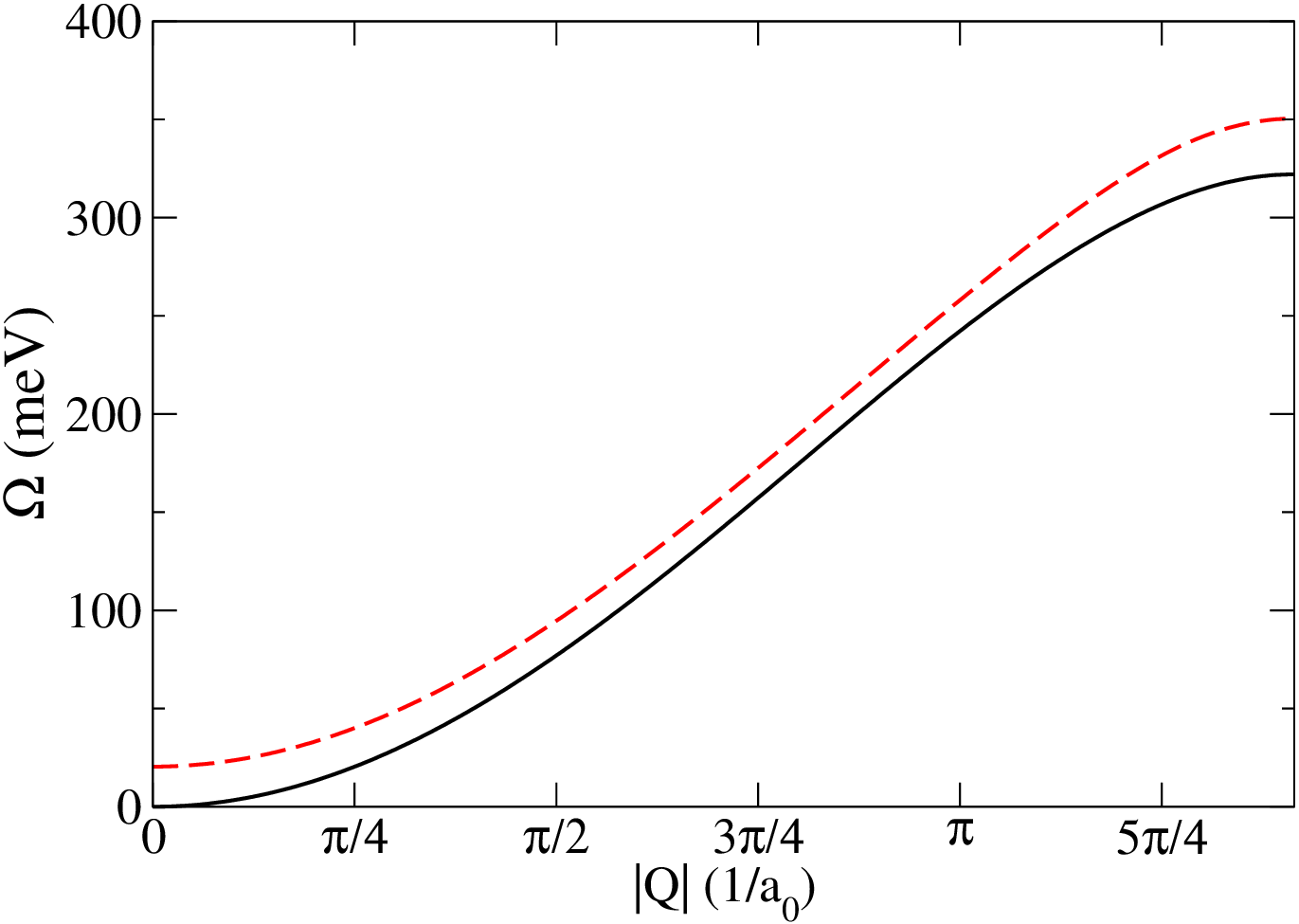}}
\caption{The dispersion relation of the two lowest lying modes in the
Co film on Cu(100). The wave vector is directed along the [11]
direction in the two dimensional Brillouin zone.}
\label{fig6}
\end{figure}

In Fig. \ref{fig5}, the left hand figure shows the eigenvector of the lowest
mode at zero wave vector (open circles), and at a reduced wave vector
of 0.6 (triangles). In the Heisenberg model, the eigenvectors are
real, it should be remarked. We see that once again the lowest mode
becomes localized onto the interface layer with increasing wave
vector. The right hand panel shows the second mode at the center of
the Brillouin zone (open circles) and its behavior at the reduced wave
vector of 0.6 (triangles). We see behavior very similar to that found
for Fe on Au(100). It is also the case here, for instance, that the
lowest mode is localized more strongly to the interface than the
second mode is to the surface. In Fig. \ref{fig6}, we show the dispersion curve
calculated for these two modes, along the [11] direction in the
zone. Despite the very different character of their eigenvectors, the
splitting in frequency of the two modes is rather small throughout the
zone. Thus, except near the zone center it is difficult for us to
resolve these two modes in the full dynamical calculations.

\section{Concluding Remarks}

We have presented studies of the nature of the spin wave modes, and
the wave vector dependence of their damping for the two low lying
modes of Fe films on the Au(100) surface. The mode excited in FMR, the
uniform mode at the center of the two dimensional Brillouin zone,
evolves into the mode localized at the interface between the film and
the substrate with increasing wave vector. The next highest mode in
frequency, a standing wave spin wave at the center of the Brillouin
zone, evolves into a surface spin wave with increasing wave vector. We
remark that while there is an extensive literature on the nature of
surface spin waves in Heisenberg magnets \cite{7}, we are unaware of any
other study of surface or interface spin waves within the framework of
a discussion that employs a realistic electronic structure and an
itinerant electron description of the ferromagnet. We do wish to point
out Mathon’s very interesting studies of surface spin waves on the
(100) surface of a one band Hubbard model \cite{20}. In his first paper, he
makes explicit contact with properties of surface spin waves generated
from Heisenberg models, and in the second paper a discussion is given
in terms of adiabatically calculated exchange integrals, generated
from the itinerant electron picture.  To return to the results
presented here, we see also that with increasing wave vector, the wave
vector dependent contribution to the damping rate of the lowest mode
in the film increases as $Q^4$. We find very similar behavior for the two
lowest lying spin wave modes for the eight layer Co film on Cu(100),
though since the localization phenomenon takes place farther out in
the Brillouin zone, we have had to resort to a Heisenberg model
description of these waves.  These two examples for films with a very
different structure suggests to us that this behavior may be expected
for other systems as well.

There are two implications of the results discussed above. First, in
the SPEELS study of the spin waves for Co on Cu(100), Vollmer et
al. suggested that their spectra received its dominant contribution
from the lowest lying spin wave mode of the film. The results here
suggest this is not so, since the lowest mode appears to localize at
the film/substrate interface with increasing wave vector, with the
consequence that its amplitude in the surface layer samples by the
SPEELS electrons is in fact very small. It is the second mode which
appears to localize on the outer surface. Thus, if one wishes to
interpret the spectra in terms of a single mode, the lowest lying mode
of the film is not the correct choice. Of course, have emphasized
earlier \cite{2,3,4,5}, except rather near to the zone center, the damping of
the spin wave modes in these ultrathin films is sufficiently strong
that it is difficult to assign the single very broad structure found
in the spectral density to a selected mode.

A second implication follows from the wave vector dependence we find
for the linewidth. In ref. \onlinecite{1}, where finite wavelength spin waves
were excited in an ultathin ferromagnet, it was argued that the data
indicates the damping to be strongly wave vector dependent. These
authors argued that two magnon scattering \cite{21,22} was responsible
for the wave vector dependence inferred from this data. The films in
ref. \onlinecite{1} were grown on exchange biased substrates. We note that in
earlier work, direct measurements of the wave vector dependence of the
spin wave linewidth in such samples were reported and found to be in
remarkable agreement with the theory of two magnon damping \cite{23}. We
show here that there is also a strong wave vector dependence to the
spin pumping contribution to the linewidth of long wavelength spin
wave modes in the itinerant ferromagnetic films. However, because of
the $Q^4$ variation we have found, the effect is small until one reaches
wave vectors in the range of $10^7$~cm$^{-1}$. Thus, for the purposes of the
data in ref. \onlinecite{1}, which explore much longer wavelength modes, one can
regard the spin pumping contribution to the wave vector dependence of
the damping rate to be quite negligible. Our results thus reinforce
the interpretation offered in ref. \onlinecite{1}.

\acknowledgments{
The research of D. L. M. was supported by the U. S. Department of
Eneergy, through grant no. DE-FG03-84ER-45083. A. T. C. and
R. B. M. have received support for the CNPq, Brazil. A. T. C. also
acknowledges the use of computational facilities of the Laboratory for
Scientific Computation, UFLA, Brazil.
}

\end{document}